\documentclass[aps,pra,twocolumn,groupedaddress]{revtex4}
\usepackage{graphicx}
\usepackage{dcolumn}

\begin{document}

\title{Quantum Particles Constrained on Cylindrical Surfaces with Non-constant Diameter}

\author{Nobuhisa Fujita}
\email[]{nobuhisa@struc.su.se}

\affiliation{Department of Structural Chemistry, Arrhenius Laboratory,
Stockholm University, 10691 Stockholm, Sweden}

\date{\today}

\begin{abstract}

We present a theoretical formulation of the one-electron problem constrained
on the surface of a cylindrical tubule with varying diameter.
Because of the cylindrical symmetry, we may reduce the problem to a
one-dimensional equation for each angular momentum quantum number $m$ along
the cylindrical axis. The geometrical properties of the surface determine the
electronic structures through the geometry dependent term in the equation.
Magnetic fields parallel to the axis can readily be incorporated.
Our formulation is applied to simple examples such as the catenoid and the sinusoidal
tubules. The existence of bound states as well as the band structures, which are
induced geometrically, for these surfaces are shown.
To show that the electronic structures can be altered significantly by applying
a magnetic field, Aharonov-Bohm effects in these examples are demonstrated.

\end{abstract}


\keywords{quantum particles, constraint, curved surfaces, curvature, cylindrical symmetry}

\maketitle

\section{Introduction}

Developments in the synthesis of nano-structured materials is ever expanding
the list of structures which may be used for applications.
Among other things, structures obtained as self-assembled interfaces in
lyotropic liquid crystals, which are used as the templates of meso-porous
materials, as well as structures made of bent graphites, including carbon
nano-fullerenes, are intriguing examples. 
These structures are characterized by curved surfaces in nanometer scales and
often possess a variety of geometrical properties. 
It is interesting to study the electronic properties related to the geometrical
properties if the electron is constrained on the curved surfaces, because
such unique settings may provide a new phenomena in quantum domain which can be
pursued for electronic applications.
We here consider the problem of quantum particle motion constrained on
curved surfaces with the tubular topology and cylindrical symmetry.

We start with a model approach developed in Refs.\cite{jensen,costa,ikegami},
in which the electron is assumed to be confined strongly to within a thin layer
built over the surface.
A limiting procedure that brings the thickness of the layer to be zero leads
to the Hamiltonian in the form of the sum of a Laplace-Beltrami operator and an
effective attractive potential energy associated with the local curvatures.
These terms are written using double curvilinear coordinates on the surface.
The stationary solutions of the Schr$\ddot{\rm o}$dinger equation are
inevitably affected by the metric properties, curvatures, and topology of the
surface.
Previous studies reported that the effective potential energy can induce bound
states whose localization centers are strongly curved parts of the surface.
\cite{costa,ikegami}
The discrete energy levels for bound states usually lie below the lowest edge
of the essential (continuous) spectrum.
The electronic structures can also be affected substantially by the global
topology of the surface, because it restricts the ways in which the
electronic wave propagates and interferes. This point, however, has not been
studied comprehensively.

The purpose of this paper is to develop a basic formulation for curved
surfaces with the tubular topology and cylindrical symmetry, and also to 
present some of its outcomes for simple examples.
Our attention is focused on the case of tubular surfaces with non-constant
diameter, namely, ones with non-vanishing Gaussian curvatures.
The geometrical character of the surfaces dealt with in this paper is among
the simplest and can be treated with relatively simple mathematics, though
the results are not trivial.
Within our formulation, it can be readily shown that there can be bound states
lying inside the essential spectrum with a different angular momentum.
The analyses of the electronic structures of curved surfaces with more
complicated geometries, especially ones so called triply periodic minimal
surfaces,\cite{aoki} should resort to more elaborated group theoretical
techniques. Our formulation is expected to provide some insight into further
studies.

This paper is organized as follows. In Section 2, the basic
Schr$\ddot{\rm o}$dinger equation for an electron strongly constrained on a
curved surface is summarized briefly.\cite{jensen,costa,ikegami}
The equation is applied to a class of tubular surfaces
with cylindrical symmetry in Section 3. By symmetry, the basic two-dimensional
equation can be reduced to a one-dimensional form, where an axial
magnetic field can also be taken into account. In Section 4 we study several
specific examples and
provide the functional form of the potential energy for each case.
These specific problems are solved numerically,
where we observe the existence of bound states at the strongly curved region on
a surface. For periodically modulated structures, the relevant
one-dimensional band structures as well as charge density distributions along
the tubule are calculated.
The electronic structures can be manipulated by tuning the applied
magnetic field. The Aharonov-Bohm oscillations of the energy levels of the
bound states are demonstrated.
Further discussions are given in Section 5, and the conclusions in Section 6.

\section{One-particle motion constrained on a curved surface}
\label{sec:formltn}

Consider a two-dimensional manifold $\mathcal{M}$ embedded in the
three-dimensional Euclidean space $E^3$ and assume a curvilinear surface
coordinate system $q^{i}$ ($i=1,2$) on $\mathcal{M}$.
Then the metric tensor $g_{ij}$ ($i,j=1,2$) is defined by
${\rm d}s^2=g_{ij}{\rm d}q^{i}{\rm d}q^{j}$ where ${\rm d}s$ is an
infinitesimal distance.
We write the principal curvatures as $\kappa_i$ ($i=1,2$).
The quantum mechanics of a single electron which is constrained physically
on $\mathcal{M}$ through a strong confining potential energy is described
by the following Schr$\ddot{\rm o}$dinger equation,
\cite{jensen,costa,ikegami}
\begin{eqnarray}\label{SchEq}
-\frac{\hbar^2}{2M}\frac{1}{\sqrt{g}}\frac{\partial}{\partial q^i}
\sqrt{g}g^{ij}\frac{\partial}{\partial q^j}\psi(q^1,q^2)&&\nonumber\\
-\frac{\hbar^2}{8M}(\kappa_1-\kappa_2)^2\psi(q^1,q^2)&=&E\psi(q^1,q^2),
\end{eqnarray}
where $g=\det{(g_{ij})}$ and $(g^{ij})=(g_{ij})^{-1}$.
The first term involves the Laplace-Beltrami operator which describes the
propagation of waves through a curved surface with non-Euclidean metric.
The second term is the attractive potential energy associated with the
local curvature,\cite{vsclassical} which essentially arises from the 'spread'
of the wave function perpendicular to the surface and thus is related to the
uncertainty principle.
The normalization condition for $\psi(q^1,q^2)$ is given by the integral
\begin{eqnarray}
\int\int |\psi(q^1,q^2)|^2 g^{\frac{1}{2}} {\rm d} q^1 {\rm d} q^2,
\end{eqnarray}
which should be unity for square integrable cases.
Static magnetic fields can also be taken into account, in which case the equation becomes
\begin{eqnarray}\label{SchEqWithB}
\frac{1}{2M}\frac{1}{\sqrt{g}}\left(
-i\hbar\frac{\partial}{\partial q^i}+\frac{e}{c}A_i \right)
\sqrt{g}g^{ij}\left(
-i\hbar\frac{\partial}{\partial q^j}+\frac{e}{c}A_j \right)\nonumber\\
\times \psi(q^1,q^2) -\frac{\hbar^2}{8M}(\kappa_1-\kappa_2)^2 \psi(q^1,q^2)=E\psi(q^1,q^2),
\end{eqnarray}
where $A_i$ are the components of the magnetic vector potential associated
with the coordinates $q^i$.

\section{Surfaces with Cylindrical Symmetry}

Now let us consider a class of smoothly curved surfaces with cylindrical
symmetry, so that they are invariant under an arbitrary rotation around
the $z$ axis.
It is convenient to use the cylindrical coordinates $(r,\phi,z)$,
with which the generator for a rotation around the $z$-axis is given by
$l_z=-i\partial/\partial \phi$, the $z$-component of the angular momentum.
Any of such surfaces can be specified by the set of coordinates $(z(x),r(x))$
as functions of the longitudinal surface coordinate $x$ taken
perpendicular to the azimuthal coordinate, $\phi$. A given value of $x$
thus specifies a circle of radius $r(x)$ lying on $z=z(x)$ plane.
If the function $z(x)$ has a monotonic dependence so that it is invertible,
then the radius can be expressed as a function of $z$, $r=r(x(z))=r(z)$.
This means that, in describing such a surface, we may use the $z$ coordinate
as the longitudinal surface coordinate.
We shall treat this particular case in more detail in the following.

For the cylindrical surface specified as $r=r(z)$, the surface metric ${\rm d} s^2=g_{zz} {\rm d} z^2 +g_{\phi\phi}{\rm d} \phi^2$ is given by
\begin{eqnarray}
g_{zz}&=&1+{r^{\prime}(z)}^2,\nonumber\\
g_{\phi\phi}&=&r(z)^2,\\
g\quad &=&r(z)^2(1+{r^\prime(z)}^2).\nonumber
\end{eqnarray}
The principal curvatures of the surface are given by
\begin{equation}
\kappa_1=\frac{1}{r(z)\sqrt{1+{r^{\prime}(z)}^2}}, \quad
\kappa_2=-\frac{r^{\prime\prime}(z)}{(1+{r^\prime(z)}^2)^{\frac{3}{2}}}, 
\end{equation}
where $\kappa_2$ corresponds to the curvature along the longitudinal
direction.

We shall consider a magnetic field which is applied to the system parallel to
the $z$-axis, and suppose its magnitude is given by $B(r)$, a function of the
radial coordinate $r$.
In the symmetric gauge, the vector potential is given by
\begin{equation}
(A_x,A_y,A_z)=\frac{\Phi(r)}{2 \pi}\frac{1}{r^2}(-y,x,0),
\end{equation}
where $\Phi(r)=\int^r_0 2\pi r^{\prime}B(r^{\prime}){\rm d} r^{\prime}$ is the total
flux passing through a disc of radius $r$.
In the cylindrical coordinates $(r,\phi,z)$, the relevant components of the
vector potential are given by
\begin{equation}
(A_r, A_\phi, A_z)=(0, \frac{\Phi(r)}{2 \pi}, 0).
\end{equation}
This can be applied for an Aharonov-Bohm type magnetic field,
$A_\phi=\Phi/2\pi$, as well as a uniform magnetic field, $A_\phi=B r^2/2$.
The Schr$\ddot{\rm o}$dinger equation can now be written as
\begin{eqnarray}
-\frac{\hbar^2}{2M}\left[\frac{1}{r\sqrt{1+{r^\prime}^2}}\frac{\partial}{\partial z}\frac{r}{\sqrt{1+{r^\prime}^2}}\frac{\partial}{\partial z}-\frac{1}{r^2}\left(l_z+\frac{e}{c\hbar}A_{\phi}\right)^2\right]\nonumber\\
\times \psi(z,\phi) -\frac{\hbar^2}{8M}(\kappa_1-\kappa_2)^2\psi(z,\phi)=E\psi(z,\phi),
\end{eqnarray}
where $l_z=-i\frac{\partial}{\partial \phi}$.
Since $l_z$ commutes with the Hamiltonian (the rotational symmetry), we can assume
as the form of the solution $\psi(z,\phi)=e^{im\phi}\psi_m(\phi)$, in which
$m$ is the eigenvalue of $l_z$ and can take only integer values in order for the
wave function to be single valued.
Then the Schr$\ddot{\rm o}$dinger equation is written as
\begin{eqnarray}
-\frac{\hbar^2}{2M}
\left[\frac{1}{r\sqrt{1+{r^\prime}^2}}
\frac{\rm d}{\rm d z}\frac{r}{\sqrt{1+{r^\prime}^2}}
\frac{\rm d}{\rm d z}-\frac{\lambda_m(z)^2}{r^2}
\right]\psi_m(z)\nonumber\\
-\frac{\hbar^2}{8M}(\kappa_1-\kappa_2)^2\psi_m(z)=E\psi_m(z),
\label{equation}
\end{eqnarray}
where
\begin{equation}
\lambda_m(z) \equiv m+ \frac{e}{c\hbar}A_\phi(r(z)).
\end{equation}
This has a Sturm-Liouville form,
\begin{eqnarray}
-\frac{\rm d}{{\rm d} z}\left[p(z)\frac{{\rm d}\psi_m(z)}{{\rm d} z}\right] + q(z)\psi_m(z) =\epsilon w(z) \psi_m(z), 
\end{eqnarray}
where
\begin{eqnarray}
\epsilon=\frac{2ME}{\hbar^2},
\end{eqnarray}
and
\begin{eqnarray}
p(z)&=&\frac{r}{\sqrt{1+{r^{\prime}}^2}},\nonumber\\
q(z)&=&\left[\frac{\lambda_m^2}{r^2}-\frac{1}{4}(\kappa_1-\kappa_2)^2\right]w(z),\\
w(z)&=&r\sqrt{1+{r^{\prime}}^2}.\nonumber
\end{eqnarray}
Hence by introducing the new variable $x$ as
\begin{equation}
x=\int^z \sqrt{\frac{w(z^{\prime})}{p(z^{\prime})}}{\rm d}z^{\prime}=\int^z \sqrt{1+{r^{\prime}(z^{\prime})}^2}{\rm d}z^{\prime},
\label{variablex}
\end{equation}
Eq.(\ref{equation}) can be reduced to a simpler form,\cite{CourantHilbert}
\begin{equation}\label{SLtr1}
-\frac{{\rm d}^2}{{\rm d}x^2}y_m(x)+W_m(x)y_m(x)=\epsilon y_m(x),
\end{equation}
where 
\begin{equation}
y_m(x)=(p(z)w(z))^{\frac{1}{4}}\psi_m(z)=\sqrt{r(z)}\psi_m(z),
\end{equation}
and
\begin{eqnarray}
W_m(x)&=&\frac{q(z)}{w(z)}+\frac{1}{(p(z)w(z))^{\frac{1}{4}}}\frac{{\rm d}^2}{{\rm d}x^2} (p(z)w(z))^{\frac{1}{4}}\label{Wm1}\\
&=&\left(\lambda_m(z)^2-\frac{1}{4}\right)\frac{1}{r(z)^2}-\frac{1}{4}\kappa_2^2.\label{Wm2}
\end{eqnarray}
Eq.(\ref{SLtr1}) has the form of a one-dimensional Schr$\ddot{\rm o}$dinger
equation with a potential energy $W_m(x)$, where $x$ corresponds to the
Euclidean line length along a curve on the surface with fixed $\phi$.
Note that the second term in Eq.(\ref{Wm1}) turns out to be
$(\kappa_1^2-2\kappa_1\kappa_2-1/r^2)/4$,
so that all terms including $\kappa_1$ are formally cancelled out in the
last expression (\ref{Wm2}).
The normalization condition for the wave function $y_m(x)$ is
\begin{equation}
2\pi\int|y_m(x)|^2{\rm d} x=1,
\end{equation}
for square integrable cases.


\section{Specific examples}

We shall study several particular examples of curved surfaces
with cylindrical symmetry by using the formulation presented above.
The first and the simplest example is the catenoid, which gives
one of the simplest examples of curved surfaces with zero mean curvature or
{\it minimal surfaces}.
The second example is the tubular surfaces whose diameter changes periodically
along the cylindrical axis, in which the diameter function $r(z)$ is given
by a sinusoidal form.

For each of these simple surfaces, the Schr$\ddot{\rm o}$dinger equation is
solved numerically, using a finite difference method with appropriate boundary
conditions.
For instance, a Dirichlet or Neumann boundary condition
(see Appendix \ref{boundary}) can be used for the case of the catenoid, while
the constant phase increment for Bloch states is used for the case of
the sinusoidal surface.

\subsection{Catenoid}

\begin{figure}[hbt]
\begin{center}\includegraphics[width = 5.5cm]{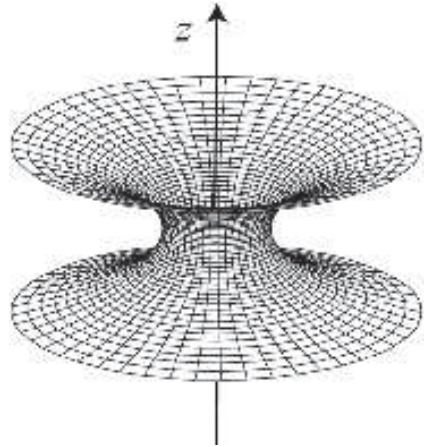}
\caption{The catenoid surface defined by $r(z)=a\cosh{z}$ with $a=1$.
The value of $a$ corresponds to the inner radius of the pore in unit of
the characteristic pore depth along $z$.}\label{fig1}
\end{center}\end{figure}

The catenoid is defined in its generalized form by the radius function
$r(z) = a \cosh{z}$,
in which $a$ ($>0$) specifies the aspect ratio. The surface is minimal if and
only if $a=1$ (see Fig.\ref{fig1}), corresponding to the original definition
of the catenoid.
Such surfaces can be thought of as simple geometrical models of pore
structures common in physical systems. Note that the
unit of length is taken as the characteristic depth of the pore.\cite{pore}

The variable $x$ defined in Eq.(\ref{variablex}) is given by
\begin{equation}
x = \int_0^z \sqrt{1+a^2 \sinh^2 z^\prime} {\rm d} z^\prime = -i E(iz,a^2),
\label{catenoid_x(z)}
\end{equation}
where the function $E(\phi,k^2)=\int^\phi_0 \sqrt{1-k^2 \sin^2 \phi} {\rm d} \phi$ is the elliptic integral of the second kind.
The potential energy function reads
\begin{equation}
W_m(x)=\left(\lambda_m^2-\frac{1}{4}\right)\frac{1}{a^2 \cosh^2 z}
-\frac{1}{4}\frac{a^2 \cosh^2 z}{(1+a^2 \sinh^2 z)^3}
\end{equation}
where $z$ implies a function of $x$. Note that in this expression the
'radial' part and the 'longitudinal' part compete with each other as $a$
is varied.
The radial part, which is dominant when $a$ is small, decays outside the
pore region of unit depth. Without magnetic field, this contribution is
negative (attractive) for $m=0$ while it is positive (repulsive) for $m\ne 0$.
This is plausible because in a narrow part of the catenoid the
contribution of kinetic energy due to alternation of signs of a wave
function around the axis becomes costly unless the wave function is
effectively repelled from that region. This part also inherits some attractive
contribution associated with the curvatures.
On the other hand, the longitudinal part is dominant when $a$ is large, and
it decays rapidly when $x$ exceeds $1/a$. This is purely the effect of the
curvature in the longitudinal direction and independent of $m$ as well
as the magnetic field.

\begin{figure}[b]
\begin{center}\includegraphics[width = 8.5cm]{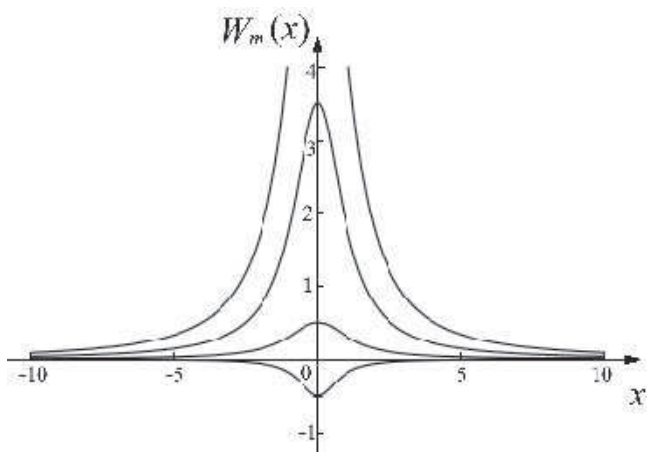}
\caption{Effective potential energy curves $W_m(x)$ for catenoid when there
is no magnetic field.
These four curves corresponds to $m=0$ (bottom), $m=\pm 1$, $m=\pm 2$,
and $m=\pm 3$ (top),
respectively.
}\label{fig2}
\end{center}\end{figure}

In the case of $a=1$, the above expressions reduce to
\begin{equation}
x = \int_0^z \cosh z {\rm d} z = \sinh z
\end{equation}
and
\begin{equation}
W_m(x)=\left(\lambda_m^2-\frac{1}{4}\right)\frac{1}{1+x^2}
-\frac{1}{4}\frac{1}{(1+x^2)^2}.
\end{equation}
This is the simplest form among the examples we have considered apart from a
trivial straight tubule.\cite{exact?} Several potential energy curves $W_m(x)$
for $a=1$ without magnetic field are shown in Fig.\ref{fig2}.
The minimal value of the potential energy for $m=0$, $W_0(0)=-1/4(1/a^2+a^2)$,
is maximal for $a=1$, so that the two contributions can be said to be
balanced for this case.


\begin{figure}[t]
\begin{center}\includegraphics[width = 8.0cm]{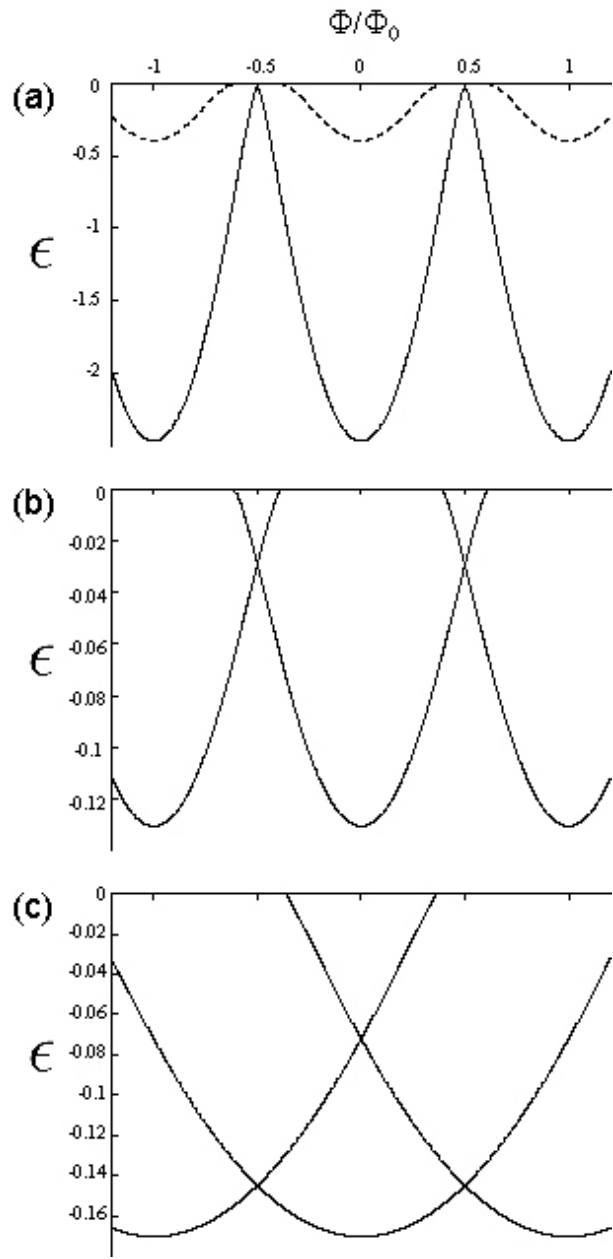}
\caption{The dependence of the eigenenergies of the bound states
on the Aharonov-Bohm flux through the pore of the catenoid, for (a) $a=0.25$,
(b) $a=1.0$, and (c) $a=2.5$. Three congruent curves (shifted horizontally)
correspond to $m=-1$, $0$, and $1$. For $a=0.25$, there exists the second
bound state for each $m$ and is shown by a broken line.}
\label{fig3}
\end{center}\end{figure}

We have numerically obtained the energy spectrum for the catenoid surfaces
with several different values of $a$, where in most cases we have found
one or several bound states centered around the constricted part of the
surfaces.\cite{boundstate}
The energies of the bound states lie below the continuum $E>0$. For instance,
one bound state is found for $a=1.0$ without magnetic field and the energy
eigenvalue has been numerically estimated to $\epsilon\approx -0.13$.

For an Aharonov-Bohm type magnetic field, the magnetic field is non-zero
only along the rotational axis, so that the surface is totally field-free.
Hence, any solution of Eq.(\ref{SchEqWithB}) can be given by a solution of
the field free equation, (\ref{SchEq}) multiplied by the phase factor
\begin{equation}
\exp{\left(-\frac{ie}{\hbar c}\int^q_{q_0}\vec{A}\cdot {\rm d}\vec{l}\right)},
\end{equation}
which will affect the boundary condition.
In the case of cylindrical surfaces, this factor is given by
\begin{equation}
\exp{\left(-i\frac{e\Phi}{h c}\phi\right)},
\end{equation}
which is independent of $z$.
Note that the function $\lambda_m(z)$ in Eq.(\ref{equation}) is a constant
\begin{equation}
\lambda_m(z)\equiv \lambda_m = m+\frac{e\Phi}{hc},
\end{equation}
and the energy spectrum is the superposition of the sub-spectra for all the
allowed values of $\lambda$'s.
This indicates that the energy spectrum oscillates as a function of flux
$\Phi$ (Fig.\ref{fig3}) and the period of oscillation is given by:
\begin{equation}
\Phi_0=\frac{hc}{e}\approx 2.07\times 10^{-11} {\rm Gauss}\cdot m^2.
\end{equation}
In Fig.\ref{fig3} we show the energy eigenvalues of bound states as functions
of the magnetic flux for $a=0.25$, $1.0$, and $2.5$.
The energy eigenvalue is minimal when the total flux is zero and it increases
rapidly with the magnetic flux. We can also see that the energy spectrum is
less sensitive for larger $a$ consistent with the fact that the longitudinal
part of $W_m(x)$ is field independent.
In the case of uniform magnetic field the oscillation is suppressed for the
magnetic flux passing through the tubule is no longer a constant along the
axis direction if the radius is not constant.

\subsection{Sinusoidal tubules}

\begin{figure}[hbt]
\begin{center}\includegraphics[width = 8.5cm]{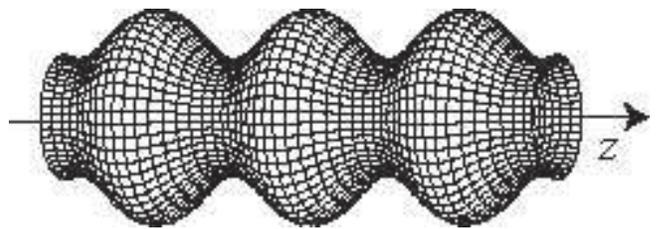}
\caption{A sinusoidal surface defined by $r(z)=a+b\cos z$ with $a=1.5$ and $b=0.5$.}
\label{fig4}
\end{center}\end{figure}

\begin{figure}[hbt]
\begin{center}\includegraphics[width = 8.5cm]{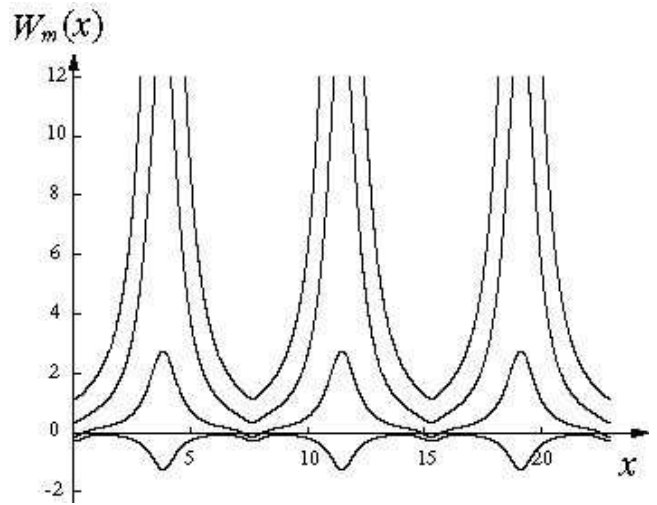}
\caption{Effective potential energy curves $W_m(x)$ for the sinusoidal surface ($a=1.5$, $b=1.0$) when there is no magnetic field.
These four curves corresponds to $m=0$ (bottom), $m=\pm 1$, $m=\pm 2$,
and $m=\pm 3$ (top), respectively.}
\label{fig5}
\end{center}\end{figure}

\begin{figure}[hbt]
\begin{center}\includegraphics[width = 8.5cm]{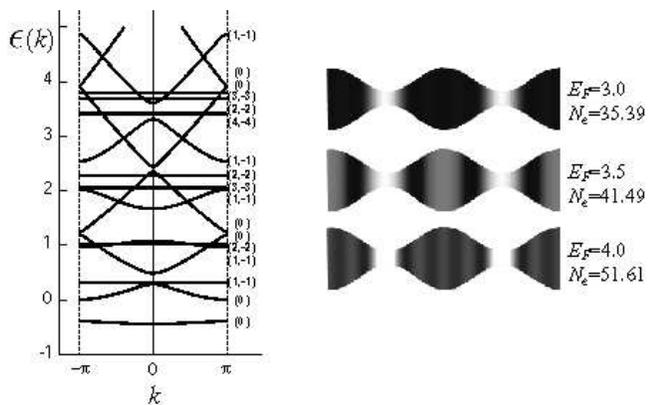}
\caption{(left) The energy band structures for a sinusoidal surface with
$a=1.5$ and $b=1.0$ without magnetic field, in which the values of $m$
corresponding to each band is shown in parenthesis.
Note that for fixed $m$, energy gaps are the most prominent in a lowest
part of the spectrum, and it becomes dominant for larger $m$.
(right) Charge density distributions along the sinusoidal surface with three
different Fermi energies $E_F=3.0$, $3.5$, and $4.0$, corresponding to
the number of electrons per period $N_e=35.39$, $41.49$, and $51.61$,
respectively. The brighter color corresponds to high concentration of
the charge density.}
\label{fig6}
\end{center}\end{figure}

\begin{figure}[hbt]
\begin{center}\includegraphics[width = 8.5cm]{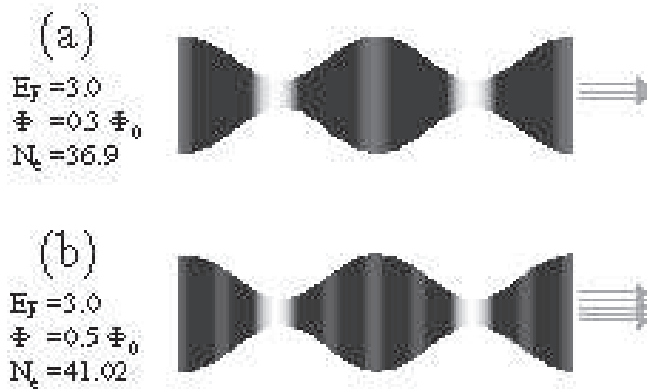}
\caption{The charge density distribution calculated for $E_F=3.0$ with two different values of magnetic flux, (a) $\Phi=0.3\Phi_0$ and (b) $\Phi=0.5\Phi_0$.}\label{fig7}
\end{center}\end{figure}

We next consider the case of infinitely many pores along the tubule.
This is achieved by taking a radius function of the sinusoidal form
$r(z)=a+b\cos (z)$ with $a>b$ ($>0$), so that the diameter changes
periodically along the axis. (See Fig.\ref{fig4}.)
The relevant variable $x$ is given by
\begin{equation}
x=\int^z_0 \sqrt{1 + b^2 \sin^2 z^{\prime}} {\rm d} z^{\prime} = E(z,-b^2).
\end{equation}
Then the effective potential energy curve,
\begin{equation}
W_m(x)=\left(\lambda_m^2-\frac{1}{4}\right)\frac{1}{(a+b\cos z)^2}
-\frac{1}{4}\frac{b^2 \cos^2 z}{(1 + b^2 \sin^2 z)^3},
\end{equation}
will be periodic in $x$ (see Fig.\ref{fig5}).
The solutions of the Schr$\ddot{\rm o}$dinger equation for fixed $m$ conform
to Bloch's theorem, resulting in the energy bands which depend on geometrical
parameters of the tubule. The total energy spectrum is the superposition of
sub-spectra for all $m$'s. The consequences of the modulating diameter of
the system is evident from potential energy curves.
Those for the case of constant diameter (straight tubules), these functions
take merely constant values depending on $m$, posing a constant bias
shifts for the energy spectrum.  
In the case of a tubule with modulating diameter, the effect of modulation
(energy gaps) is prominent in the lowest part of the spectrum for each $m$.
The effect of modulation can also be seen in the charge density distribution
profiles shown in Fig.\ref{fig6} with several different Fermi energies.
Due to the Aharonov-Bohm oscillation of the electronic structures, the
charge density distribution also changes for different values of magnetic
flux enclosed in the tubule. (Fig.\ref{fig7})

\section{Discussions}

The basic equation (Sec.\ref{sec:formltn}) assumes hypothetically that the
surface has zero thickness and the potential energy within the surface is flat.
One should bear in mind that these assumptions pose essential requirement for
the scale of Fermi wavelength ($\lambda_F$); that is, it should be
sufficiently larger than the surface thickness and the average atomic
distance along the surface.
This is necessary in order for the motion of an electron perpendicular to the
surface to be strongly quantized and for the flat potential approximation
to be valid.
The condition would be satisfied if a constrained electronic system is
realized using semiconductor hetero-structures ($\lambda_F\sim 100$nm)
or bent carbon graphite sheets\cite{VT92} ($\lambda_F\sim 10$nm).
On the other hand, the model should be regarded as an excessive
oversimplification for curved surfaces made of metallic substances,
in which $\lambda_F\sim 0.1$nm.

Here we assume the scale of a pore diameter to be of the order of $a\sim 5$nm,
which is typically the scale of pore structures in mesoporous materials and
some of carbon nanotubes.
Then the magnetic field should be as large as
$B\sim\frac{\Phi_0}{a^2}=\frac{hc}{ea^2}\approx10^6
{\rm Gauss}=10^2{\rm Tesla}$ in order to make the total flux passing through
the pore to be $\Phi_0$. This is rather large since the largest static field
available is about 50 Tesla.
The electronic properties could nevertheless change significantly with more
moderate field strength if gap openings and/or closings may be induced by
the magnetic field. It might be interesting to point out here that measurements
of the Aharonov-Bohm effect in carbon nanotubes have been achieved recently.
\cite{ABoscNT1,ABoscNT2,ABoscNT3}

In our formulation, the initial Schr$\ddot{\rm o}$dinger equation reduces to
the one-dimensional equations each of which corresponds to an angular momentum
quantum number $m$ along the cylindrical axis.
The effective potential energy $W_m(x)$ reflects the geometry of the tubule
as well as the applied field, so that we may actually create a wide range
of one-dimensional potentials by tuning these factors.
Since the spectra for different $m$'s are independent of each other,
a bound state for a $m$ can exist within the range of the essential
spectrum for another $m$. We have found a simple example with this property
in the case of a straight tubule with inflated part, $r(z)=a+b/(1+x^2)$
($a,b>0$).
It is important to note that the response to a change of geometry as well
as to an applied field differs among different $m$ states, while the Fermi
energy can cross several partial energy spectra for different $m$'s.
One may therefore pursue similar systems for applications such as electronic
resonators or a selective STM probe for different $m$ states. Note that
the possibility of using carbon nanotubes as STM probes has been reported.
\cite{STMprobe1,STMprobe2}

If we combine multiple tubules together, more profound and interesting
effects would be expected due to the interaction between neighboring
tubules. It is also interesting to study more complicated surfaces
like triply periodic curved surfaces\cite{aoki} which are often encountered
in soft matter systems as well as mesoporous materials.
In such studies, one may focus on the inherent
effects of the surface metric, curvatures, symmetry and topology, where our
present study may also serve as preliminary information.
We may also pursue the effects of interparticle correlation, which should
be essential to understand real systems.

\section{Conclusions}

We have formulated the problem of quantum particle motion
constrained on curved tubular surfaces with cylindrical symmetry.
As the surface has the continuous rotational symmetry, the two surface
coordinates can be separated and the problem is finally reduced to a
one-dimensional differential equation.
Magnetic fields parallel to the axis is also taken into account in the
formulation. Several examples are studied numerically.

\section*{Acknowledgments}

The author would like to thank O. Terasaki for his suggestion to investigate
similar problems.
He also thanks A. Laptev and T. Yokosawa for helpful discussions, and
T. Nakajima for sending useful information.

\appendix

\section{Finite difference method and boundary conditions}\label{boundary}

Using the finite difference method, the original equation (\ref{SLtr1})
can be written as
\begin{eqnarray}
-\frac{1}{a^2}y((j-1)a)+\frac{2}{a^2}y(ja)
-\frac{1}{a^2}y((j+1)a)\nonumber\\
+W(ja)y(ja)=\epsilon y(ja),
\end{eqnarray}
where $a$ is the mesh size.
Several different boundary conditions have been used to obtain numerical
solutions of the one-dimensional Schr$\ddot{\rm o}$dinger equation.
First, the Dirichlet boundary conditions can be used for the case of
non-periodic surfaces in which the effective one-dimensional potential energy
$W_m(x)$ behaves asymptotically as $W_m(x)\to 0^{-}$ ($x\to\pm\infty$).
The influence of the cutoff $\pm X$ of space variable $x$ becomes
exponentially small for a bound state which is localized in a sufficiently
smaller region than $X$, while it will decrease as $\propto 1/X$ for a
scattering state. In both cases, one can estimate the energy spectrum fairly
well using this boundary condition with sufficiently large $X$. If the
boundary is far enough from the center, so that $|W_m(X)|\ll|\epsilon|$,
we can improve the precision of the result by taking the Neumann boundary
conditions instead. This means we assume the form $y(X+a)=y(X)e^{-\kappa a}$
for $\epsilon=-\kappa^2<0$ and $y(X+a)=y(X)e^{-ika}$ for $\epsilon=k^2>0$,
where the value of $\epsilon$ should be determined self-consistently.
For the case of periodic sinusoidal surfaces, we may use the
usual assumption of Bloch states, which will lead to a closed form of matrix
equation. This can be used to obtain the band structures.




\begin{thebibliography}{50}
\bibitem{vsclassical} Since the curvature potential at any given point is determined not only by the intrinsic curvatures but also by the extrinsic ones, the problem is essentially distinct from the classical counterpart, as discussed in Ref. 2).
\bibitem{pore} It can be taken as the pore (inner) radius, in which case $r(z)=\cosh(az)$ and the following equations must be scaled appropriately.
\end{thebibliography}
\end{document}